\documentclass[12pt]{article}

\usepackage[scale={.83,.83}]{geometry}
\usepackage{amsmath,amssymb,amsfonts,amsthm,dsfont,dcolumn}
\usepackage[margin=-5pt]{caption}
\usepackage{graphicx,wrapfig}
\usepackage{tensor}
\usepackage{empheq}
\usepackage[hypertex]{hyperref}
\usepackage{latexsym}
\usepackage{yfonts}

\newcolumntype{d}{D{.}{.}{0}}
\newlength{\textlength}
\newlength{\overlinelength}

\numberwithin{equation}{section}

\begin{document}

\newcommand{\be}{\begin{equation}}
\newcommand{\ee}{\end{equation}}
\newcommand{\bea}{\begin{eqnarray}}
\newcommand{\eea}{\end{eqnarray}}
\newcommand{\beqn}{\begin{eqnarray}}
\newcommand{\eeqn}{\end{eqnarray}}
\newcommand{\ack}[1]{[{\bf Pfft!: #1}]}

\newcommand{\eref}[1]{eq.\ (\ref{eq:#1})}
\def\NPB{{\it Nucl. Phys. }{\bf B}}
\def\PL{{\it Phys. Lett. }}
\def\PRL{{\it Phys. Rev. Lett. }}
\def\PRD{{\it Phys. Rev. }{\bf D}}
\def\CQG{{\it Class. Quantum Grav. }}
\def\JMP{{\it J. Math. Phys. }}
\def\SJNP{{\it Sov. J. Nucl. Phys. }}
\def\SPJ{{\it Sov. Phys. J. }}
\def\JETPL{{\it JETP Lett. }}
\def\TMP{{\it Theor. Math. Phys. }}
\def\IJMPA{{\it Int. J. Mod. Phys. }{\bf A}}
\def\MPL{{\it Mod. Phys. Lett. }}
\def\CMP{{\it Commun. Math. Phys. }}
\def\AP{{\it Ann. Phys. }}
\def\PR{{\it Phys. Rep. }}

\hyphenation{Min-kow-ski}
\hyphenation{cosmo-logical}
\hyphenation{holo-graphy}
\hyphenation{super-symmetry}
\hyphenation{super-symmetric}

\def\pa{\partial}
\newcommand{\de}{\mathtt{z}} 
\newcommand{\beq}{\begin{equation}}
\newcommand{\eeq}{\end{equation}}
\newcommand{\ftt}{{f_{tt}(T)}}

\rightline{VPI-IPNAS-10-01}
\rightline{ MCTP-10-02}
\centerline{\Large \bf}\vskip0.25cm
\centerline{\Large \bf }\vskip0.25cm
\centerline{\Large \bf Aging and Holography}
\vskip .5cm

\centerline{\bf
Juan I. Jottar\footnote{jjottar2@illinois.edu},
Robert G. Leigh\footnote{rgleigh@illinois.edu},
Djordje Minic\footnote{dminic@vt.edu} and
Leopoldo A.  Pando Zayas\footnote{lpandoz@umich.edu}
}
\vskip .5cm

\centerline{\it ${}^{1,2}$Department of Physics}
\centerline{\it University of Illinois}
\centerline{\it 1110 W. Green Street}
\centerline{\it Urbana, IL 61801, U.S.A.}
\vskip 0.5cm
\centerline{\it ${}^3$Institute for Particle, Nuclear and Astronomical Sciences}
\centerline{\it Department of Physics, Virginia Tech}
\centerline{\it Blacksburg, VA 24061, U.S.A.}
\vskip 0.5cm
\centerline{\it ${}^4$Michigan Center for Theoretical Physics}
\centerline{\it Randall Laboratory of Physics}
\centerline{\it University of Michigan}
\centerline{\it Ann Arbor, MI 48109, U.S.A.}
\vskip .5cm

\begin{abstract}
Aging phenomena are examples of `non-equilibrium criticality' and can be exemplified by
systems with Galilean and scaling symmetries but no time translation invariance. We realize aging
holographically using a deformation of a non-relativistic version of gauge/gravity
duality. Correlation functions of scalar operators are computed using holographic real-time techniques, and agree
with field theory expectations. At least in this setup, general aging phenomena are
reproduced holographically by complexifying the bulk space-time geometry, even in Lorentzian signature.

\end{abstract}


\newpage

\section{Introduction}

Gauge/gravity duality has been used extensively to study both quantum critical behavior
and equilibrium and hydrodynamical properties of a wide range of strongly-coupled field
theories. Time-dependent phenomena are typically much more challenging.
Such phenomena are of significant interest both on the field theory side, as well as from
the purely gravitational point of view. Understanding systems that are out of equilibrium
is one of the most challenging problems in physics today, and it certainly would be of
interest to see if gauge/gravity duality can shed some light on this subject.

In this paper, we will make a first step in this direction. One of the simplest notions of
non-equilibrium physics is {\it aging}, which corresponds  to `non-equilibrium
criticality,' in a sense that we will make more precise below. If a physical system is
brought rapidly out of equilibrium by a sudden change of an external control parameter
(quenching), one often finds (i) slow (non-exponential) dynamics, (ii) breaking of time
translation invariance and (iii) dynamical scaling \cite{HenkelPleimlingBook}, \cite{Henkel:2007bc}. We will show that
aging phenomena \cite{Minic:2008xa} can be studied by deforming a variant of gauge/gravity
duality \cite{Son:2008ye,Balasubramanian:2008dm}
with isometries given by the Schr\"odinger algebra. The geometrization of the Schr\"odinger algebra was first studied long ago; see for example, \cite{Duval:1984cj,Duval:1990hj} and the more recent \cite{Duval:2008jg}. For a comprehensive review of the subject of aging from the condensed matter physics perspective, see
\cite{Bray94}, \cite{Cugl02} for example.

A crucial step in the theoretical description of some far from equilibrium phenomena was
the realization of the role of local scale invariance
\cite{Henkel:2007bc,Henkel:1993sg,Henkel:2002vd,Henkel:2006kt}. In the particular case of aging phenomena
\cite{michelbook}, where it is observed that the properties of non-equilibrium systems
generically depend on time since the system was brought out of equilibrium, the role
of a sub-algebra of the Schr\"odinger algebra with dynamical exponent $\de=2$ has been found to
be crucial \cite{Henkel:2003wt,Henkel:2004pc}. This algebra is called the aging algebra, $Age_d$.

In the recent past, there has been much interest in applying gauge/gravity duality to condensed matter systems. The
classic examples of gauge/gravity duality involve anti-de Sitter (or asymptotically anti-de Sitter) space-times, which are dual to (states of) conformal field theories. There are also versions of gauge/gravity duality that possess non-relativistic, rather than relativistic, symmetries,
the most symmetric being the Schr\"odinger algebra which is generated by the Galilean symmetries as well as scale transformations. It is this algebra that is the closest non-relativistic version of the relativistic conformal algebra that plays a central role in AdS/CFT duality. Work
on this includes \cite{Son:2008ye,Balasubramanian:2008dm,Volovich:2009yh,Leigh:2009ck,Leigh:2009eb} as well as finite temperature black hole versions \cite{Rangamani:2008gi,Herzog:2008wg,Maldacena:2008wh,Adams:2008wt}. 

In this paper, we explore aging in the context of gauge/gravity duality. Our principle results include the construction of time-dependent
gravitational backgrounds that are dual to $Age_d$-invariant field theories. The paper is organized as follows. In Section \ref{sec:Schrodinger} we review the structure of the Schr\"odinger algebra $Schr_d$ and correlation functions in Schr\"odinger invariant field theories. We
also review how these theories are studied holographically. On the gravity side, $Schr_d$ is realized as the isometry algebra acting on fields propagating on the background space-time, while on the dual field theory side there is a distinct representation of $Schr_d$ acting on the operators of the theory. In Section \ref{sec:aging}, we review some features of the aging algebra and how it is represented, drawing on and in some cases, interpreting, the vast literature. In Section \ref{Sec:Metrics}, we present our first main results concerning the construction of geometries with $Age_d$ isometries. The principle result there is the realization that such geometries have dimension $d+3$ (as do the standard geometries with $Schr_d$ isometry) and locally have $Schr_d$ isometry. In order to bring these geometries into the standard $Schr_d$-invariant form, a singular time-dependent transformation is required, and the locations of these singularities are to be thought of as the temporal locations in which the dual field theory is kicked out of equilibrium. In Section \ref{sec:correlators} we consider the holographic computation of correlation functions in the $Age_d$-invariant field theories, using appropriate real-time techniques. We find that the correlation functions so obtained are of the form expected for $Age_d$-invariant field theories. There is a very interesting subtlety that emerges in this comparison however: the most general aging phenomena are reproduced by {\it complexified metrics} (even in Lorentzian signature). We believe that this is sensible: the complexity is related directly to the non-equilibrium character of the system, in particular its `decay' towards an equilbrium configuration.  We conclude in Section \ref{sec:conclusions}.

\section{The Schr\"{o}dinger Algebra and Representations}\label{sec:Schrodinger}

The Schr\"odinger group is an action on space and time coordinates extending the usual
Galilean symmetries to include anisotropic scaling $\vec x\to\lambda \vec x, t\to
\lambda^\de t$. Throughout this paper, we will only discuss the non-relativistic case,
$\de=2$. These actions close on a larger group generated by temporal translations $H$,
spatial translations $P_{i}$, Galilean boosts $K_{i}$, rotations $M_{ij}$, dilatations $D$
and the special `conformal' transformation $C$.

The commutation relations satisfied by these generators, aside from the obvious
commutators of rotation and translations and those that vanish, read

\begin{align}\label{eq:Schrdalg}
\left[D,K_{i}\right] &= K_{i},& \left[D,P_{i}\right] &= -P_{i},& \left[P_{i},K_{j}\right] &= \delta_{ij}N\\
 \left[D,C\right] &= 2C,&  \left[C,P_{i}\right] &= -K_{i},&  \left[H, C\right] &= D\\
 \left[H,K_{i}\right] &= P_{i},& \left[D,H\right] &= -2H\label{eq:Schrdalg3}
\end{align}
We note that $\{H,D,C\}$ generate an $SL(2,\mathds{R})$ subalgebra, with $\{P_i,K_i\}$ forming an $SL(2,\mathds{R})$ doublet, and the generator $N$ is central.
We will refer to the full algebra as $Schr_d$, where $d$ is the spatial dimension. This algebra made its first appearance long ago, for example, as the invariance group of the Schr\"odinger equation with zero potential.

If one considers field theories with this symmetry, one finds, in complete analogy with
the conformal field theory bootstrap, \cite{Belavin:1984vu} that correlation functions are
of a restricted form \cite{Henkel:2007bc}. For example, for operators that are scalars under rotations,
the two-point function is essentially given by 
\be
\langle {\cal O}_1(t_1, \vec{x}_1){\cal O}_2(t_2, \vec{x}_2) \rangle= \delta_{\Delta_1, \Delta_2} \delta_{n_1+n_2,0}
(t_{1}-t_{2})^{-\Delta_1} \exp\left(i\frac{n_2}{2} \frac{(\vec{x}_{1}-\vec x_{2})^{\,2}}{t_{1}-t_{2}}\right)\, .
\ee
In the holographic context, this result
was obtained using real-time methods \cite{Skenderis:2008dh,Skenderis:2008dg}  in Ref. \cite{Leigh:2009eb}, and using Euclidean methods in \cite{Fuertes:2009ex,Volovich:2009yh}. Similarly, higher point functions are of a constrained form as well.

%

In holography, one makes use of a space-time possessing $Schr_d$ as its algebra of
isometries. Such a space-time may be taken to have
metric \cite{Son:2008ye,Balasubramanian:2008dm}
\begin{equation}\label{Son-McGreevy}
ds^{2} = \frac{L^{2}}{z^{2}}\left(dz^{2} - \frac{\beta^{2}}{z^{2}}dt^{2} - 2dtd\xi + d\vec{x}^{2}\right),
\end{equation}
where $\vec{x} = (x_{1},\ldots, x_{d})$ are the spatial coordinates of the dual field theory and
$z$ is the holographic direction (with the ``boundary" located at $z = 0$). The parameters
$\beta$ and $L$ are length scales, and all the coordinates have units of length.

The space-time \eqref{Son-McGreevy} has Killing vectors
\beqn\label{general Killing vector nu one}
M_{ij}&=& x_{j}\partial_{i} - x_{i}\partial_{j}\\
P_{i} &=& \partial_{i},\ \ \ H= \partial_{t}, \ \ \ \ N= \partial_{\xi}\\
C &=& zt\,\partial_{z} + t^{2}\,\partial_{t} + \frac{1}{2}\left(z^{2} + \vec x^{2}\right)\partial_{\xi} + tx^{i}\,\partial_{i}\\
D&=& z\,\partial_{z} + 2t\,\partial_{t} + x^{i}\partial_{i}\\
K_{i} &=&  t\,\partial_{i} + x^{i}\partial_{\xi} 
\eeqn
providing a representation of the $Schr_d$ algebra (\ref{eq:Schrdalg}--\ref{eq:Schrdalg3}) acting on bulk (scalar) fields. Since $N$ is central, such fields can be taken to be equivariant with respect to $N$, i.e., their $\xi$-dependence can be taken to be of the form $e^{in\xi}$ for fixed\footnote{$\xi$ is often taken to be compact, so that the spectrum of operators in the dual theory is discrete. This is not without problems in the bulk, as $\pa_\xi$ is null in the geometry (\ref{Son-McGreevy}).} $n\in\mathds{R}$.  In gauge/gravity
duality, the asymptotic ($z\to 0$) values of fields propagating in this geometry act as sources for
operators in the dual field theory. The $Schr_d$ algebra acts on those operators in a way
that can be deduced from the field asymptotics \cite{Leigh:2009eb,Leigh:2009ck}. We thus
get another distinct representation of $Schr_d$ that acts on (scalar) operators of the dual field theory
\beqn
M_{ij}&=& x_{j}\partial_{i} - x_{i}\partial_{j}\\
P_{i} &=& \partial_{i},\ \ \ H= \partial_{t}, \ \ \ \ N= \partial_{\xi}\\
D&=& 2t\,\partial_{t} + x\partial_{x} + \Delta\, \mathds{1}\\
C &=&  t^{2}\,\partial_{t}  + tx\,\partial_{x} + \frac{x^{2}}{2}\,\partial_{\xi}+ \Delta t\, \mathds{1}\\
K_i&=&  t\,\partial_{i} + x_i\partial_{\xi} 
\eeqn
where $\Delta$ is the scaling dimension; for equivariant fields, $\pa_\xi$ evaluates to $in$.

\section{The Aging Algebra and Correlation Functions}\label{sec:aging}

The aging algebra, which we will denote as $Age_d$, is obtained by discarding the time
translation generator $H$ from the $Schr_d$ algebra. Indeed, the form of the algebra is
such that it makes sense to do so. This is the simplest possible notion of time-dependent
dynamics, and it is considered as a rather special form of non-equilibrium physics. 
In this paper, we consider the problem of constructing an appropriate space-time geometry possessing $Age_d$ as its isometry algebra, and then compute some simple correlation functions.

 Since $H$ has been discarded, such correlation functions are not generically time-translation invariant. To see what sort of time-dependence to expect, let us consider a construction that often appears in the literature.
Consider a diffusive system with (white) noise
$\eta$ and a time-dependent potential $v(t)$ governed by a wave-function $\phi$ satisfying\footnote{The
diffusion equation should be regarded as a Wick rotated version of a Schr\"odinger-like
equation (or equivalently, a Schr\"odinger-like equation is obtained by considering $M$ to
be imaginary). In following sections, we will always work in Lorentzian signature. One does not expect that such Wick rotations are innocuous in general.}
\be
\label{dyneqn}
2 M \partial_t \phi = \nabla^2 \phi - \frac{\delta V}{\delta \phi} - v(t) \phi + \eta.
\ee
As pointed out in \cite{HenkelPleimlingBook}, correlation functions in this system can be studied by examining deterministic
dynamics governed by the aging group. Note that the `gauge transformation'
\be
\label{GaugeTransformation}
\phi \to \phi \exp\left(-\frac{1}{2M} \int^t v(\tau) d \tau\right),
\ee
removes the time dependent potential term from equation (\ref{dyneqn}).
This means that this sort of time-dependence can be mapped to a system governed by   the Schr\"odinger group. A simple physical model
for this type of time dependence is the out of equilibrium decay of a system towards equilbrium, following some sort of quench.

In the special case
\be
v(t) \sim 2M Kt^{-1},
\ee
one finds that the wavefunctions are related via a scaling function to the wavefunctions of the Schr\"odinger problem
\be
\label{aging-schrodinger}
\phi_{Age} = t^K \phi_{Schr}.
\ee
%
Thus, the local scale-invariance of aging systems is largely determined by studying first
the Schr\"odinger fields. In particular, the correlators of operators in an $Age_d$-invariant theory can be expressed in terms of  the correlators of operators in a $Schr_d$-invariant theory. Schematically, for the two-point function of a scalar operator $\mathcal{O}$, the result we want to reproduce is \cite{Picone:2004id},\cite{HenkelPleimlingBook}

\begin{equation}\label{schematic expected form of correlator}
\langle \mathcal{O}(t_{1})\mathcal{O}(t_{2})\rangle_{Age} \sim \left(\frac{t_{1}}{t_{2}}\right)^{\#}\langle \mathcal{O}(t_{1})\mathcal{O}(t_{2})\rangle_{Schr},
\end{equation}

\noindent where $\#$ is a constant which characterizes the breaking of time-translation invariance. We will present the details of this relationship below.
%
%
%

\subsection{Representations of $Age_d$}\label{sec:agerep}

Above, we gave two representations of $Schr_d$, one acting on (scalar) operators of a field theory,
and one acting on the holographic bulk (scalar) fields.
We now consider removing $H$ from the algebra, and ask how the representation of the
remaining generators might be modified. We will consider first the representation on operators, as that is what appears 
in the literature. Let us assume that the  time and spatial coordinates
$t$ and $\vec x$ and the non-relativistic mass (equivalently the coordinate $\xi$) retain their
standard meaning. Consequently, we take the representation of $M_{ij},P_i,N,K_i$ and $D$ to be unchanged from that of $Schr_d$.
However, it is possible that the form of the generator $C$ could be modified consistent with the $Age_d$ algebra. Suppose then that we write
$C_{A} = C + \delta C$, where $C_{A}$ is the
representation of $C$ in the aging algebra. In order for $C_{A}$ to satisfy the
commutators of the $Age_d$ algebra (which are unchanged from $Schr_d$) we need
\begin{equation}
\left[P_i,\delta C\right]=0, \qquad \left[N,\delta C\right]=0, \qquad \left[K_i,\delta C\right]=0, \qquad \left[D,\delta C\right] = 2\delta C.
\end{equation}
The first three commutators are easily seen to imply that $\delta C$ can only be of the form
$\delta C = g_{1}(t)\mathds{1} + g_{2}(t)\partial_{\xi}$. The fourth commutator then fixes $g_{i}(t)$ ($i=1,2$):
\begin{equation}
\left[D,\delta C\right] = 2\delta C \qquad \Rightarrow \qquad t\partial_{t}g_{i}(t) = g_{i}(t) \qquad \Rightarrow \qquad g_{i}(t) = K_{i}t,
\end{equation}
and we then conclude \cite{Picone:2004id} that the most general form is
\begin{equation}\label{C in Aging}
C_{A} =  t^{2}\,\partial_{t}  + tx\,\partial_{x} + \frac{x^{2}}{2}\,\partial_{\xi} + (\Delta + K_1) t\, \mathds{1} + K_2t\,\partial_{\xi},
\end{equation}
where $K_1$ and $K_2$ are constants. Note that in the undeformed case $K_1=K_2=0$, and that when evaluated on equivariant fields, $K_2$ is accompanied by the eigenvalue of $\partial_\xi$ which is imaginary. Thus, $\delta C = Kt\,\mathds{1}$, where $K=K_1+inK_2$ is naturally a complex number. The physical significance of the real and imaginary parts will be discussed more fully below, but for now we note that we expect $K$ to make an appearance in time non-translation invariant features of correlation functions. 

$K$ has been described in the literature \cite{Picone:2004id} as a `quantum number' labeling representations. We believe that it is better to think of this parameter as a property of the {\it algebra} instead; indeed, later we will see $K$ emerge in a holographic setup as a parameter appearing in the metric, rather than being associated with any particular field. We further note that if we were to demand $\left[\pa_t,C_{A}\right] =D$, we  would find that $K=0$, thus recovering the full Schr\"odinger algebra in its
standard representation.


\section{A Geometric Realization of the Aging Group }\label{Sec:Metrics}

In this section, we will explore how we might implement $Age_d$ as an isometry algebra.
There are a variety of possible constructions that present themselves. First, we might consider the breaking
of time translation invariance to be associated with the introduction of a `temporal defect'. In relativistic gauge/gravity duality, there is a way to introduce spatial defects \cite{DeWolfe:2001pq,Aharony:2003qf} by placing a D-brane along an $AdS_{d-1}$ slice of $AdS_d$. Such a brane intersects the boundary along a co-dimension one curve, which can be coordinatized as $x=0$. The choice of slicing preserves as much symmetry as possible: clearly $P_x$ is broken along with $K_x$ and $M_{xj}$,\footnote{Here we refer to the generators of the $AdS_d$ isometry, $SO(d,2)$.} leaving unbroken $SO(d-1,2)$. Because of Lorentz invariance, presumably such a construction works for temporal defects in the relativistic case (see \cite{Bak:2007qw} for related work on time-dependent holography). Such a construction would involve placing an S-brane along a suitable slice.

Similarly, spatial defects \cite{Karch:2009rj} can be introduced in non-relativistic holography in much the same way. One can imagine placing a D-brane along a $Schr_{d-1}$ slice of the $Schr_d$ space-time. Here, $P_x$, $K_x$ and $M_{x,j}$ would be broken. It is fairly obvious though that in this case, a temporal defect cannot be constructed in this way. The basic reason is that time is much different than space in a non-relativistic theory, and at the algebraic level, we seek to break $H$ only. We will see signs of this sort of difficulty below.

\subsection{Coset construction of aging metrics}\label{sec:coset}
To investigate the possibility of finding space-times that admit $Age_d$ as their isometry, we consider
first a coset construction. This approach allows us to explore the precise realization of certain symmetries
linearly or non-linearly and it is an exhaustive way of understanding the dimensionality of the space-time in which the aging algebra can act. Ultimately, we will conclude that metrics with $Age_d$ isometry are
realized in the {\it same dimension} as $Schr_d$. Along the way, one will appreciate the distinct difference
between temporal and spatial defects.


The goal of this section is to construct a $G$-invariant metric on the coset space $M=G/H$. The classic example \cite{Nappi:1993ie} is Mink$_d$=Poincar\'e$_d$/Lorentz$_d$, in which $P_\mu$ are realized linearly, while the other generators are realized non-linearly. We notice in particular that $\dim M=\dim G-\dim H$, and at the algebraic level, the generators split into two sets ${\cal H}$ and ${\cal M}$.

We give a few details of the construction here, but the interested reader should consult the literature. We write generators generically as $T_n$, and use the notation $T_{[m]}$ to denote generators in the coset, and $T_h$ to denote generators in $H$. We introduce structure constants $f$ and in particular are interested in the commutators $[T_h,T_{[n]}]=f_{[n] h}^{\ [j]} T_{[j]}$. The sought-after metric of $M$ is related to a symmetric bilinear form $\Omega_{mn}$ in the generators $T_n$ which is non-degenerate and invariant, satisfying:
\beq
\label{bilinear}
\Omega_{[m][n]}f_{[k]p}^{[m]}+\Omega_{[k][m]}f_{[n]p}^{[m]}=0\, .
\eeq

First, let us review how one might obtain the $Schr_d$ space-time in this way \cite{SchaferNameki:2009xr}. One takes $G$ to be the Schr\"odinger group, and identifies $H$ as generated by ${\cal H}=\{M_{ij}, K_i,C-\frac12\gamma N\}$ with $\gamma$ a real (non-zero) parameter, while ${\cal M}=\{H,D,N,P_i\}$. Counting parameters, we see that this will give rise to a space-time of dimension $d+3$. The structure constants of interest may be deduced from the $Schr_d$ commutators given above
\beqn
f_{[H] K_j}^{[P_i]}=f_{[P_i] K_j}^{[N]}=-\delta_{ij},\ \ \ \ f_{[D] (C+\gamma N)}^{[N]}=-\gamma,\ \ \ \ \
f_{[H] (C+\gamma N)}^{[D]}=-1,\ \ \ \ \
f_{[P_k] M_{ij}}^{[P_\ell]}=\delta_{ik}\delta_{j\ell}-\delta_{jk}\delta_{i\ell}
\eeqn
Eq. (\ref{bilinear}) then gives the Killing metric (using the ordering $\{H,D,N,P_i\}$)
\beq
\Omega=\begin{pmatrix} -b & 0 & -a & 0\cr 0 & \gamma a & 0 & 0\cr -a & 0 & 0 & 0\cr 0 & 0& 0&a \end{pmatrix}
\eeq
Notice that this is non-degenerate only when $\gamma\neq 0$.
If we represent a group element as $g=e^{x_H H} e^{x_N N}e^{x^i P_i}e^{x_D D} $,
we compute\footnote{Here, we just need the relation $e^{-xD}Ae^{xD}=e^{-nx}A$ if $[D,A]=nA$.}
\beq
g^{-1}dg = e^{2x_D} dx_H H+e^{x_D} dx^i P_i+dx_N N+dx_D D.
\eeq
We then find
\beq
\Omega(g^{-1}dg,g^{-1}dg)= -be^{4x_D}dx_H^2+ae^{2x_D}(-2dx_Ndx_H+dx_i^2)+\gamma adx_D^2.
\eeq
This is precisely the Schr\"odinger metric in the form (\ref{Son-McGreevy}), if we identify $e^{x_D}=L/z, x_H=t, x_N=\xi$ and $\gamma=L^2, a=1, b=\beta^2/L^2$.

Next, we try to apply a coset construction to derive a space-time with $Age_d$ isometry. If we simply take the coset above, but remove $H$ from ${\cal M}$, we would construct a space-time of dimension $d+2$. However, one finds that there is no such coset with non-degenerate metric; the operational reason is that if we eliminate the structure constants with an $[H]$ index, then the relation (\ref{bilinear}) gives $\Omega_{[N][any]}=0$. We conclude that there is no coset construction of $Age_d$ isometric space-times in dimension $d+2$. This means that there is no way to slice $Schr_d$ in co-dimension one that preserves $Age_d$ along the slice, and thus there is no analogue of a `defect construction' for a temporal defect in this non-relativistic case. As shown in Ref. \cite{Karch:2009rj}, one can do this for spatial defects, and in the context of the coset construction, it just means that we can construct a space-time with $Schr_{d-1}$ isometry in $d+2$ dimensions.

We can attempt to construct $Age_d$ in $d+3$ dimensions via the coset construction. In this case, we could take
\beq
{\cal H}=\{M_{ij},P_i\},\ \ \ \ {\cal M}=\{C,D,N,K_i\}
\eeq
and obtain the structure constants of interest
\beq
{f_{[C]}}_{P_j}^{[K_i]}=-\delta_{ij},\ \ \ \ \
{f_{[K_i]}}_{P_j}^{[N]}=-\delta_{ij},\ \ \ \ \
{f_{[K_k]}}_{M_{ij}}^{[K_\ell]}=\delta_{ik}\delta_{j\ell}-\delta_{jk}\delta_{i\ell}
\eeq
Working out the Killing metric, which has to satisfy (\ref{bilinear}), we find
\beq
\Omega=\begin{pmatrix} -a & -c & -w & 0\cr -c & b & 0 & 0\cr -w & 0 & 0 & 0\cr 0 & 0& 0&w \end{pmatrix},
\eeq
and we note $\det\Omega = -bw^{d+2}$. Representing a group element as
$
g=e^{x_c C} e^{x_N N}e^{x^i K_i}e^{x_D D}\, ,
$
we compute
\beq
g^{-1}dg = e^{-2x_D} dx_C C+e^{-x_D} dx^i K_i+dx_N N+dx_D D \, ,
\eeq
and hence
\beq
\Omega(g^{-1}dg,g^{-1}dg)= -e^{-4x_D}adx_C^2+bdx_D^2-2ce^{-2x_D}dx_Cdx_D+we^{-2x_D}dx_i^2-2we^{-2x_D}dx_Ndx_C \, .
\eeq
We will hold off on interpreting the coordinates that appear here until the next section. However, we note that this metric does not actually have $Age_d$ as its isometry, but in fact $Schr_d$, as long as all of the generators are well-defined.

\subsection{Direct Construction of $Age_d$ Isometric Space-times}

We have learned that in attempting to construct space-times with $Age_d$ isometry, we are lead to space-times that have $Schr_d$ isometry. We believe the precise statement is that the space-times constructed in this way are {\it locally Schr\"odinger}. That is, at a generic point, the isometry algebra is $Schr_d$. This leaves open the possibility that there might exist space-times that are locally Schr\"odinger, but globally have only $Age_d$. A simple example of how this might occur is that given a metric with $Age_d$ but no manifest time translation invariance, a {\it singular} time-dependent change of coordinates is necessary to bring the metric into the standard Schr\"odinger form. This possibility, which we will realize below, is not in conflict with the coset construction given above. Now, one might argue that such coordinate singularities are not physical, but this is far from the truth in holographic studies. The coordinate singularities that we will study below are on a similar footing in this regard to black hole horizons. These are of course only coordinate singularities with no invariants behaving poorly, but there is no doubt that they have a profound effect in gauge/gravity duality --- the horizon is the place where we impose boundary conditions.

In this section, we explore a more direct approach to constructing metrics with $Age_d$ isometry. That is, we consider a class of metrics that manifestly preserve the generators of $Age_d$ but are not time-translation invariant. Our main result, mentioned above, is that a metric admitting
the algebra of aging as isometries automatically admits an extra generator that completes
the aging algebra back, up to an isomorphism, to the Schr\"odinger algebra. In fact, this situation reflects, on the holographic side, the close relationship between Age and
Schr\"odinger fields, as exemplified by eq.  (\ref{GaugeTransformation}). Indeed, we will go on to construct correlation functions in these geometries and show that they have the expected form.

The key feature of such metrics is that $Age_d$ invariance allows the metric components to depend on the invariant combination $T=\beta t/z^2$. This is scale invariant (and dimensionless) but clearly transforms under time translations. A fairly general Ansatz for the metric is then of the form
\begin{align}\label{metric ansatz}
ds^{2} &= \frac{L^{2}}{z^2}\left[f_{zz}\left(T\right)\,dz^2 + \frac{2\beta}{z}f_{zt}\left(T\right)\,dzdt - \frac{\beta^{2}}{z^{2}}f_{tt}\left(T\right)\,dt^{2} + f\left(T\right)\left(-2dtd\xi  + d\vec x^{2}\right)\right]\, .
\end{align}
As a regularity requirement, the functions $f_{zz}$ and $f$ are
chosen such that $\det g =-L^{8}f_{zz}f^{3}/z^{8}$ is non-zero everywhere other than
$z=\infty$. It is clear that the Ansatz \eqref{metric ansatz} already admits $\{M_{ij},P_i,K_i,D,N\}$ as
isometries and the generators are in the same form as in the Schr\"odinger space-time, (\ref{general Killing vector nu one}). If the metric functions $f_{zz}(T)$ and
$f(T)$ are independent, one concludes that there are no further isometries.

We would now like to augment the set of isometries to include a Killing vector isomorphic
to the generator $C$. We will find that an attempt to add a generator $C$ is always
accompanied, locally, by the simultaneous appearance of a generator $H$. Thus, it is not possible to
construct in this way a metric that has $Age_d$ isometry without having the full $Schr_d$
isometry, locally. As we will describe below, these Schr\"odinger metrics are a family
parameterized by 2 arbitrary functions (chosen here to be $f$ and $f_{tt}$), and a closed
expression for all the generators is given.

One finds that in order to enlarge the isometry group, we have to impose the following
relations between the metric functions:
\begin{empheq}{align}\label{fzz}
f_{zz}(T) &= \left[1 + h(T)\right]^{2}, \\
f_{zt}(T) &= T\left[f_{tt}(T) + cf^{2}(T)\right] - \frac{h(T)(h(T)+2)}{4T},\label{fzt}
\end{empheq}
where $c$ is a constant and we have defined $h(T) = T\dot{f}(T)/f(T)$ (the dot denotes
a derivative with respect to $T$). Appendix A contains more details about this construction. We note that these geometries have a constant negative scalar curvature given by
$R =-\frac{(d + 3)(d + 2)}{L^{2}}.$
Since the cosmological constant in $d+3$ dimensions is
$\Lambda= -\frac{(d + 1)(d + 2)}{2L^{2}}$,
these  metrics then satisfy (in units where $\kappa^{2} = 16\pi G_{N} =1$)
\begin{equation}\label{Einstein equation general d}
G_{\mu\nu} + \Lambda g_{\mu\nu} = R_{\mu\nu} + \frac{(d + 2)}{L^{2}}g_{\mu\nu} =\frac{1}{2}T^{(mat)}_{\mu\nu},
\end{equation}
where 
\begin{equation}
T^{(mat)}_{\mu\nu}=-2c(d + 4)\frac{\beta^{2}}{z^{4}}f(T)^{2}\,\delta^{t}_{\phantom{t}\mu}\delta^{t}_{\phantom{t}\nu} \qquad \Rightarrow \qquad \left(T^{m}\right)^{\mu}_{\phantom{\mu}\mu}=0,
\end{equation}
and $c$ is the integration constant appearing in \eqref{fzt}. An explicit system supporting the geometry (\ref{metric ansatz}--\ref{fzt}) would have to give rise to this (bulk) stress-energy tensor.

The Killing vectors are given by (the others are unchanged)
\begin{empheq}{align}\label{isometries general Schrodinger family}
C &= \frac12tz\left(\frac{h+2}{h+1}\right)\partial_{z} + t^{2}\partial_{t}   + t\vec x\cdot\vec\partial+\frac{1}{2}\left[\vec x^{2} + \frac{z^{2}}{h+1}\left(\frac{T^{2}f_{tt}}{f} + \frac{\left(h+2\right)^{2}}{4f}  +cT^{2}f\right) \right]\partial_{\xi}\\
H&= \partial_{t} + \frac{z}{2t}\frac{h}{h+1}\partial_{z} -\frac{\beta^{2}}{2z^{2}(h+1)}\left(\frac{h^{2}}{4T^2f}+ \frac{f_{tt}}{f} + c f\right)\partial_{\xi}\label{isometries general Schrodinger family 2}
\end{empheq}
and it is easy to check that they indeed satisfy the Schr\"odinger algebra. Thus, we find generators that are simply deformed from their realizations in the standard $Schr_d$ metric.

As we will explain below, it turns out that the simplified class of solutions of the form 
\beqn
f_{zt}(T)&=&T(\ftt-1)\\
f(T)&=&f_{zz}(T)=1\, ,
\eeqn
which indeed solve (\ref{fzz}-\ref{fzt}) with $c=-1$, will lead to results which match our field theoretical expectations. In this case we have
\beq\label{metric}
ds^2=\frac{L^2}{z^2}\left[ dz^2 +2\frac{\beta}{z}T(\ftt-1)dtdz -\frac{\beta^2}{z^2}\ftt dt^2-2dtd\xi+d\vec x^2\right],
\eeq
and the generators of interest simplify to
\beqn
H&=&\pa_t-\frac{\beta^2}{2z^2}(\ftt-1)\pa_\xi\\
C&=&tz\pa_z+t\vec x\cdot\vec\pa+t^2\pa_t+\frac12(\vec x^2+z^2)\pa_\xi+\frac12z^2T^2(\ftt-1))\pa_\xi
\eeqn
Clearly, this system reduces to the standard $Schr_d$ form when $\ftt=1$. It is `locally Schr\"odinger' for any $f_{tt}(T)$ in some domain. Indeed, it can be obtained from \eqref{Son-McGreevy} by the following change of coordinates:
\begin{equation}\label{transformation xi}
\xi' = \xi + \frac{\beta}{2}\int^T dT'\left[f_{tt}(T') - 1\right]\, .
\end{equation}
We comment on the properties of this coordinate transformation below. 

\subsection{Comments on the asymptotic realization of the algebra}

In gauge/gravity duality, what we are most interested in is what symmetries are realized in the dual field theory. It is clear that (\ref{transformation xi}) is singular at $t=0$ whenever $\ftt$ includes negative powers of $T$. With this in mind, we are led to consider functions of the form
\beq\label{genftt}
\ftt=\sum_j\alpha_j T^{-j}
\eeq
We will take (without loss of generality) $\alpha_0=1$ and we will find that $\alpha_1$ plays a special role. We note that $T$ contains a factor of $z^2$, which goes to zero asymptotically. Thus different powers of $T$ behave quite differently asymptotically. If positive powers of $T$ were present (i.e., $j<0$), $\ftt$ would blow up asymptotically; these would correspond to `relevant deformations' --- they correspond to terms in the generators that blow up as $z\to 0$ (for fixed generic $t$), presumably rendering, at best, some unknown algebra. Thus, we will focus on $j\geq 0$ in (\ref{genftt}).  Note then that we have
\beqn
H&=&\pa_t-\frac{\beta^2}{2}\sum_{j>0}\alpha_j (\beta t)^{-j}z^{2(j-1)}\pa_\xi,\\
C&=&tz\pa_z+t\vec x\cdot\vec\pa+t^2\pa_t+\frac12(\vec x^2+z^2)\pa_\xi+\frac12\sum_{j>0}\alpha_j (\beta t)^{2-j}z^{2(j-1)}\pa_\xi \, .
\eeqn
Terms involving $j>1$ in a sense are `irrelevant' in the  $z\to 0$ limit (for fixed generic $t$), and we thus focus on the case $\alpha_1=\alpha\neq 0$, i.e., 
\beq\label{specialftt}
f_{tt}(T) = 1 + \frac{\alpha}{T}
\eeq
which gives
\beqn
H&=&\pa_t-\frac{1}{2}\frac{\alpha\beta}{t}\pa_\xi\\
C&=&tz\pa_z+t\vec x\cdot\vec\pa+t^2\pa_t+\frac12(\vec x^2+z^2)\pa_\xi+\frac12\alpha \beta t\pa_\xi \, .
\eeqn
Notice that this particular form of $C$ is the bulk generalization of the generator we found in \eqref{C in Aging}, with $K$ identified as $i\alpha n \beta/2$. Note also that in this case, if we consider how $C$ behaves for $z\to 0$, we see that it is well-defined on the whole boundary, but $H$ is apparently singular at $t=0$. Although the full $Schr_d$ algebra is present locally, we clearly need to be careful near $t=0$. If we interpret the blowing up of the components of $H$ as an indication that we lose $H$ as a generator, we obtain precisely what we want in the dual field theory. It should be emphasized again however that this is a coordinate singularity; for example, the norm of the vector field $H$ is well-behaved everywhere in the metric\begin{align}\label{the potential age metric}
ds^{2} &= \frac{L^{2}}{z^2}\left[dz^2 + \frac{2\alpha\beta}{z}\,dzdt - \frac{\beta^{2}}{z^{2}}\left(1 + \frac{\alpha}{T}\right)\,dt^{2} -2dtd\xi  + d\vec{x}^{2}\right].
\end{align}
We note also that this metric breaks a discrete symmetry enjoyed by the $Schr_d$ metric, namely `$CT$', $t \to -t, \xi \to -\xi$.

The change of coordinates \eqref{transformation xi} that would take this metric back to the standard Schr\"odinger form is now
\begin{equation}
\label{xi-log-trans}
\xi' = \xi + \frac{\alpha\beta}{2}\ln T = \xi + \frac{\alpha\beta}{2}\ln\left(\frac{\beta t}{z^{2}}\right)\, .
\end{equation}
The multi-valuedness of the logarithm in the complex $t$ plane will play an important role in the physical interpretation of the point $t=0$ and in the calculation of correlation functions that we present below.

\subsection{Comments on coordinates}

Let us now expand on the similarity between $t=0$ and a black hole horizon. 
The point $t=0$ is a coordinate singularity --- a clear indication of this is that no curvature invariant blows up there. More transparently, the coordinate change  \eqref{transformation xi}
\be
\xi'=\xi+\frac{\beta}{2}\int^TdT'\big[f_{tt}(T')-1\big].
\ee
takes the metric into the standard Schr\"odinger form. For the particular choice of $f_{tt}=1+\alpha/T$, note the similarity between this coordinate change and the tortoise change that one performs to clarify that the horizon is a coordinate singularity in the case of the Schwarzschild solution. One can take the above coordinate change as valid away from $t=0$. Once in the coordinate $\xi'$, we can simply extend the metric past the $t=0$ singularity as we routinely do for the Schwarzschild black hole.  Thus, we should think of the above coordinate transformation as a type of tortoise coordinate that gets rid of the horizon-type singularity at $t=0$.

And yet we should ask why the analysis of Killing vectors naturally picks a metric of the form given above. As shown above our geometry is a piece of the Schr\"odinger geometry for $t>0$. We think about this piece of the geometry in much the same way as we think of the Poincar\'e patch of global $AdS$. Note that by considering the $t>0$ piece of the geometry we break time-translational symmetry, much as the Poincar\'e slice breaks rotational symmetry along the sphere of global $AdS$. Moreover, even though the Poincar\'e patch is not geodesically complete we insist on using it because we attach a specific field theoretic meaning to those coordinates. We believe that this makes sense physically and that is our goal. We found a `natural' way to cut off the Schr\"odinger space at $t=0$, which is dictated by the symmetries.

We can compare this interpretation to the treatment reviewed in Section \ref{sec:aging}. There is a special choice made in eq. (\ref{aging-schrodinger}) which relates the aging and the Schr\"odinger field. We could rewrite that equation as $\phi_{Schr}=\phi_{Age}/t^K$. So, all the aging dynamics can be written as Schr\"odinger dynamics. Of course, we throw away the point $t=0$ since in aging we actually wait for the system to start evolving a bit after the quench.

To make our analogy with the Poincar\'e patch more complete, let us consider the norm of time translation in our metric\footnote{Time translation is clearly not a symmetry of the system.}:
\be
||\partial_t||=g_{tt}=\frac{\beta^2}{z^2}f_{tt}= \frac{\beta}{z^2}\left(1+\alpha\,\,\frac{z^2}{t}\right).
\ee
For $\alpha=0$ we recall that at $z\to \infty$ the coordinate $t$ becomes null --- this is the typical horizon in the Poincar\'e patch which appears as a consequence of limiting to a patch of global $AdS$ (see, for example section 2.2.1 of \cite{Aharony:1999ti}). Now, additionally, we have that at $t\to 0$ the norm blows up.

Given the analogy with the black hole horizon, if we are to interpret the dual field theory as living in the $t>0$ patch, we must address the question of what boundary condition to place on fields at $t=0$. In the next section, we will present a calculation of two-point correlation functions. We will find, much  in the spirit of eq. (\ref{aging-schrodinger}), that solutions to equations of motion in the bulk in the $Age_d$
geometry are closely related to those in the $Schr_d$ geometry, which we note are valid for all $t$. We will take the underlying relationship with the Schr\"odinger fields as our guiding principle: $Age_d$ invariant correlation functions can be constructed via a Keldysh contour in the complex $t$-plane which passes over the point $t=0$. The behavior of fields in the vicinity of $t=0$ (namely their discontinuity) will be assumed to follow from the {\it continuity} of the corresponding Schr\"odinger fields  -- indeed, since all fields are taken to be equivariant, the coordinate change (\ref{xi-log-trans}) induces $\phi(-\epsilon,\vec{x})=e^{i\pi n\beta\alpha/2}\phi(+\epsilon,\vec{x})$. We believe that this boundary condition corresponds physically to some sort of quenching of the system, the details of which we will not further consider.

\section{Correlation Functions}\label{sec:correlators}

With these comments in mind, we now consider a scalar field on the geometry (\ref{the potential age metric}). We look for solutions to the scalar wave equation for the family of metrics \eqref{metric} that are of the form
\begin{equation}\label{new separable ansatz}
\phi(z,t,\xi, x) = e^{Q(T)}e^{in\xi}\phi_{S}(z,t,x; n) = e^{Q(T)}e^{in\xi}\int \frac{d^dp}{(2\pi)^d} \frac{d\omega}{2\pi}\, e^{i\vec p\cdot\vec x - i\omega t}\phi_{S}(\omega, p, n; z),
\end{equation}

\noindent where $Q(T)$ is some function  and $\phi_{S}$ denotes a solution of the wave equation on the standard Schr\"odinger background \eqref{Son-McGreevy} (which is translationally invariant). If such a form exists, then the appearance of time translation non-invariance in the scalar is entirely in  the scale-invariant prefactor $\exp(Q(T))$. We note that given the fact that the form of the dilatation operator is the same for both $Age_d$ and $Schr_d$, the corresponding fields have the same conformal dimension, even though there is $z$-dependence in the prefactor, precisely because $T$ is dilatation invariant. We note also that the Ansatz for the scalar fields is reminiscent of the expected form, (\ref{aging-schrodinger}).

Indeed, for the geometry (\ref{the potential age metric}), the scalar Laplacian is such that solutions are of the form (\ref{new separable ansatz}), with
\begin{equation}\label{Q of T}
Q(T)= \frac{in\beta}{2}\int^T dT'\left[f_{tt}(T') - 1\right].
\end{equation}
 For the Schr\"odinger field, in the asymptotic region we have
\beq
\phi_S(z\to 0,t,\vec x;n)\sim z^{\Delta_-} \phi^{(0)}_S(t,\vec x;n)+...
\eeq
and thus the aging field behaves (assuming the form (\ref{specialftt})) as
\beq
\phi(z\to 0,t,\vec x;n)\sim z^{\Delta_-} \left(\frac{\beta t}{z^2}\right)^{-in\beta\alpha/2}\phi^{(0)}_S(t,\vec x;n)+...
\eeq
As we mentioned, although there is an unusual factor of $z$ present here, because the generator $D$ is unmodified, the scaling dimension of $\phi$ is the same as $\phi_S$. As we will see below, the extra factor of $z$ should be thought of as a `wavefunction renormalization' factor that should be absorbed into the definition of  the dual operator --- it is included in the source for Age fields. 

Now, one might assume since the prefactor is common to both the source and the vev, that it cancels out in the evaluation of the Green function. We will show below that this is naive  --- it is related to depending too much on Euclidean methods. When one carefully examines the real-time correlation functions, one finds that the prefactor makes its presence felt.

Note also that the prefactor is generally complex. If we confine ourselves to $\alpha\in\mathds{R}$, which one normally would do in real geometry, then formally the prefactor is a phase. However, $Q(T)$ is generally a complex function, as, for the choice (\ref{specialftt}), it is a multi-valued function about $t=0$. It is then not too much of a stretch to go all the way to a complexified geometry, in the sense of taking $\alpha\in\mathds{C}$. As we have discussed above, the real and imaginary parts of $\alpha$ (or $K$ in the language of Section \ref{sec:agerep}) have a direct physical interpretation, and we will encounter precisely that in the $Age_d$-invariant correlation functions that we study below.

The Schr\"odinger theory is renormalizable \cite{Leigh:2009eb}; the on-shell action can be made finite by the inclusion of a number of {\it Schr\"odinger invariant} counterterms. One can show that the boundary renormalization of the Age theory follows that of Schr\"odinger closely. Indeed, the bulk action
\beq
S = -\frac{1}{2} \int d^{d+3}x \sqrt{-g} \left( g^{\mu\nu}\pa_\mu \bar\phi\pa_\nu \phi +
m_0^2/L^2 |\phi|^2 \right)\label{action}
\eeq
reduces on-shell to the boundary term
\begin{align}
S_{os} = \frac{1}{2}\int_\epsilon d^{d+1}x d\xi\sqrt{|\gamma|}
\ \bar\phi {\bf n}\cdot\pa\phi
\end{align}
Here we take  the `boundary' to be a constant $z$ slice, with normal $n=\frac{L}{z}dz$;  the corresponding normal vector is ${\bf n}=\frac{1}{L}(z\pa_z+\alpha\beta\pa_\xi)$ when we use $\ftt=1+\alpha/T$. Since the on-shell solutions are of the form $\phi=e^{Q(T)}\phi_{Schr}$, we get
\begin{align}\label{eq:onshellactionAge}
S_{os} = \frac{1}{2L}\int_\epsilon d^{d+1}x d\xi\sqrt{|\gamma|}
\ e^{Q(T)+\overline{Q(T)}}\bar\phi_{Schr} z\pa_z\phi_{Schr}
\end{align}
with $\gamma$ the induced spatial metric. This is precisely of the same form as in the Schr\"dinger case, with the inclusion of the prefactors appropriate to Age fields. We see again that one must be careful in this case in interpreting powers of $z$ --- since the Age fields are of dimension $\Delta$, the scale invariant $z$-dependent prefactors must be associated with the normalization of operators. They are not to be canceled by the addition of counterterms.

In this paper, we will focus on the calculation of the two-point functions of scalar operators dual to $\phi$. As we stated above, because of the time-dependence of the metric, it is dangerous to attempt to employ Euclidean continuation, and thus we will consider
the real time correlators very carefully using the Skenderis-Van Rees method \cite{Skenderis:2008dh,Skenderis:2008dg},  following \cite{Leigh:2009eb} closely.

\subsection{Review of Schr\"odinger calculations}

\begin{figure}[ht]
	\centering
	\includegraphics[height=7cm]{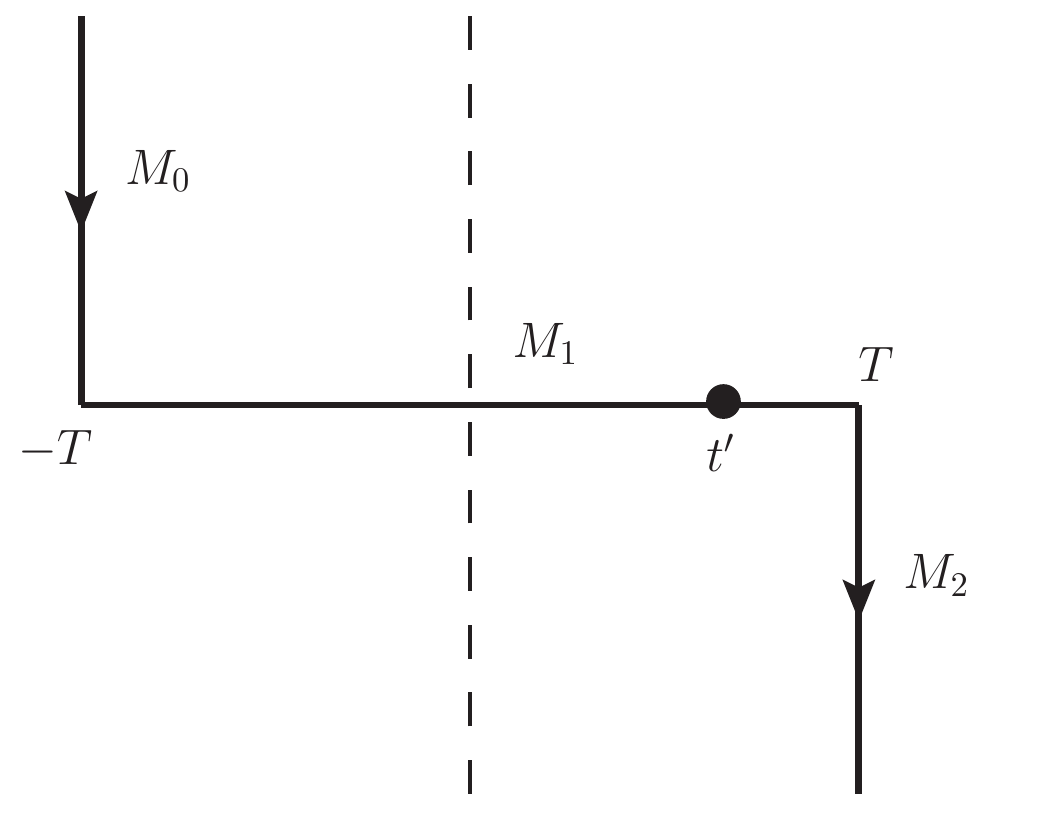}\caption{Contour in the complex $t$-plane corresponding to the time-ordered correlator.}\label{fig:Keldysh2.pdf}
\end{figure}

We refer to the reference \cite{Leigh:2009eb} for various details, but we will outline the results found there in the Schr\"odinger geometry. Generally, correlators are computed by constructing solutions along segments of a contour in the complex time plane, where the choice of contour determines the nature of the correlator considered. In the case of a time ordered correlator, the contour is as shown in Fig. \ref{fig:Keldysh2.pdf}.

A vertical segment corresponds to Euclidean signature, while a horizontal segment corresponds to Lorentzian signature. One computes the solutions in the presence of a $\delta$-function source on the real axis along each segment, and matches them at junctions.

\subsubsection{Schr\"odinger Solutions: Lorentzian}

In Lorentzian signature, we use the notation $q= \sqrt{q^2} = \sqrt{ \vec k^2-2\omega n}$. Solutions of the scalar wave equation on the Schr\"odinger space-time are given in terms of Bessel functions. For $q^2<0$, both the $K_\nu$ and $I_\nu$ solutions are regular everywhere, while for $q^2>0$, $I_\nu$ diverges for large $z$ and is discarded. Noting that $q$ has a branch point at $\omega=\vec k^2/2n$, we facilitate integration along real $\omega$ by properly deforming $q$ to $q_\epsilon=\sqrt{-2\omega n+\vec k^2-i\epsilon}$. With these comments, we arrive at  the general solution  in Lorentzian signature  \cite{Leigh:2009eb}
\beqn\label{general solu}
\phi_S(z,t,\vec x;\xi) = e^{in\xi} \int \frac{d\omega}{2\pi} \frac{d^d k}{(2\pi)^d}\
e^{-i\omega t + i \vec k\cdot  \vec x}  z^{\frac{d}{2}+1}
\left( A_s(\omega,\vec k) K_\nu(q_\epsilon z)
+ \theta(-q^2)B_s(\omega,\vec k) J_\nu(|q| z) \right)\, . 
\eeqn

\subsubsection{Schr\"odinger Solutions: Euclidean}

Next, we consider a similar analysis in Euclidean signature. To do so, we Wick rotate the metric (\ref{metric}) to (more precisely, along $M_0$ and $M_2$ respectively, we write $t=\pm(-T+i\tau)$)
\beq
ds^2=\frac{L^2}{z^2}\left[ dz^2 +\frac{\beta^2}{z^2} dt^2-2id\tau d\xi+d\vec x^2\right]\, .
\eeq
Although this metric is complex, it is possible to trace carefully through the analysis.

The general solution of the Schr\"odinger problem is
\beqn\label{E-general solu}
\phi_S(z,\tau,\vec x;\xi) &=& e^{in\xi} \int  \frac{d\omega_E}{2\pi} \frac{d^d k}{(2\pi)^d}\ e^{-i\omega_E \tau +
i \vec k\cdot  \vec x}  z^{\frac{d}{2}+1} A(\omega_E,\vec k) K_\nu(q_E z)\, ,
\eeqn
where now $q_E = \sqrt{q_E^2} = \sqrt{\vec k^2-i2\omega_E n}$. Note that in this case, the branch point is at
imaginary $\omega_E$, and so no $i\epsilon$ insertion is necessary. In writing this, we have assumed that $\tau\in(-\infty,\infty)$ and thus $\phi_S$ has no normalizable mode. One has to be careful with this. For example, if $\tau \in [0,\infty)$, we write $\omega_E = -i\omega$ for $\phi$ and
$\omega_E = i\omega$ for $\bar\phi$ and the following mode is
allowable
\beqn
\phi_S &\sim& e^{in\xi} e^{-\omega \tau + i \vec k\cdot \vec x}
z^{\frac{d}{2}+1} I_\nu(q_E z)
\eeqn
as long as $\omega > 0$ and $-2\omega n + \vec k^2 < 0$, or
equivalently $\omega > \vec k^2/2n$. For $\tau\in (-\infty,0]$, no such mode is present.

\subsubsection{Correlators (Schr\"odinger)}

The correlators are computed by recognizing the asymptotics as sources for corresponding operators
 \beq
 e^{iS^{bulk}_C[\bar\phi_S^{(0)},\phi_S^{(0)}]} = \langle
e^{i\int_C ( \hat {\cal O}^\dag\phi_S^{(0)} + \bar\phi_S^{(0)} \hat
{\cal O})}\rangle
\eeq
where the fields have asymptotic expansions
\beqn
\phi_S &=&  e^{in\xi}\Big\{ z^{\Delta_-} \left(\phi_{(0)}+ z^2 \phi_{(2)} + o(z^4)\right) + z^{\Delta_+} \left(v_{(0)} + z^2 v_{(2)} + o(z^4)\right)\Big\}\eeqn
with $\Delta_\pm=1+d/2\pm\nu$ and
\begin{align}
\phi_{(2m)} = \frac{1}{2m (2\Delta_+ - (d + 2) -
2m)}{\cal S}\phi_{(2m-2)}\, ,
\end{align}
where ${\cal S}=2in\partial_t+ \vec\nabla^2$ is the Schr\"odinger operator.
In \cite{Leigh:2009eb} the  bulk to boundary propagator was derived
\beqn
K_{(n)}(t,\vec x,z; t')&=&\frac{2 z^{1+d/2}}{\Gamma(\nu)}  \int \frac{d\omega}{2\pi} \frac{d^d k}{(2\pi)^d}\ e^{-i\omega (t-t') + i \vec k\cdot \vec x} \left(\frac{q_\epsilon}{2}\right)^\nu
K_\nu(q_\epsilon z)\, ,
\eeqn
which determines the time ordered correlator
\beq\label{time-ordered}
 \langle T\Big{(}\hat
{\cal O}_{(n)}(\vec x,t) \hat {\cal O}_{(n)}^\dag(\vec
x',t')\Big{)}\rangle_{Schr.} =
\frac{1}{\pi^{d/2}\Gamma(\nu)}\left(\frac{n}{2i}\right)^{\Delta_+-1}
\frac{\theta(t-t')}{(t- t')^{\Delta_+}} e^{in\frac{(\vec x -
\vec x')^2 + i\epsilon}{2(t-t')} }.
 \eeq
The derivation of the final result involves a somewhat difficult contour integral.

\subsection{Aging Correlators}

Now, we turn our attention to the computation of correlators holographically in the Age geometry. Because of the physical interpretation, that of a quench at $t=0$, we focus on correlators of operators inserted at times $t>0$. We have argued that the point $t=0$ is much like a horizon, and so one should in fact confine oneselves to the $t>0$ patch. As we have also discussed, the Age solutions have a discontinuity across $t=0$ given by $\phi(-\epsilon,\vec{x})=e^{i\pi n\beta\alpha/2}\phi(+\epsilon,\vec{x})$. To calculate the Age correlators, we propose that one should use the same Keldysh contour as above, fixing the solutions to the Age problem formally by requiring the continuity of the associated Schr\"odinger solutions. This prescription uniquely determines the two-point correlation function, and as we shall see, gives the expected form \eqref{schematic expected form of correlator}. We proceed to construct solutions on the segments $M_0, M_1$ and $M_2$.

\subsubsection{Age Solutions: Lorentzian}

With the same setup as for the Schr\"odinger problem above, we arrive at  the general solution  in Lorentzian signature
\beqn\label{general solu}
\phi(t,\xi,\vec x,z) = e^{Q(T)}e^{in\xi} \int \frac{d\omega}{2\pi} \frac{d^d k}{(2\pi)^d}\
e^{-i\omega t + i \vec k\cdot  \vec x}  z^{\frac{d}{2}+1}
\left( A_s(\omega,\vec k) K_\nu(q_\epsilon z)
+ \theta(-q^2)B_s(\omega,\vec k) J_\nu(|q| z) \right)
\eeqn

\subsubsection{Age Solutions: Euclidean}
\newcommand\fttE{f_{tt}(T_E)}

Next, we consider a similar analysis in Euclidean signature. To do so, we take complex time $t=\pm(-T+i\tau)$ (this is appropriate to $M_0$ and $M_2$ respectively) and thus replace the metric (\ref{metric}) by
\beq
ds^2=\frac{L^2}{z^2}\left[ dz^2 +\frac{\beta^2}{z^2}\fttE d\tau^2\mp 2id\tau d\xi\pm 2i\frac{\beta T_E}{z}(\fttE-1)d\tau dz+d\vec x^2\right]
\eeq
where $T_E=\pm\beta(-T+i\tau)/z^2$.

The general solution of the Age problem is
\beqn\label{E-general solu}
\phi(\tau,\xi,\vec x,z) &=& e^{Q(T_E)}e^{in\xi} \int  \frac{d\omega_E}{2\pi} \frac{d^d k}{(2\pi)^d}\ e^{-i\omega_E \tau +i \vec k\cdot  \vec x}  z^{\frac{d}{2}+1} A(\omega_E,\vec k) K_\nu(q_E z)
\eeqn
where now $q_E = \sqrt{q_E^2} = \sqrt{\vec k^2-i2\omega_E n}$.

\subsubsection{Time-ordered Correlator}

As in the Schr\"odinger case, the solution along the $M_0$ component is zero. Thus we have to require $\phi_1(t_1=-T,\vec x,z)=0$. We then also conclude that there is no normalizable solution on $M_1$. Thus, we should have a unique solution.

We place a $\delta$-function source at $\vec x=0, t_1=\hat t_1>0$ on $M_1$. That is, we want
\beq
\phi_1(t_1,\vec x,z)\Big|_{z\to 0}=z^{\Delta_-}e^{in\xi}\delta(t_1-\hat t_1)\delta(\vec x)
\eeq
This requires the field to be of the form
\beq\label{phi1}
\phi_{1}(t_1,\vec x,z) =\frac{2}{\Gamma(\nu)}e^{Q(T_1)-Q(\hat T_1)}e^{in\xi}  z^{1+d/2-in\alpha\beta} \int  \frac{d\omega}{2\pi} \frac{d^d k}{(2\pi)^d}\ e^{-i\omega (t_1-\hat t_1) + i \vec k\cdot \vec x}
\left(\frac{q_\epsilon}{2}\right)^\nu K_\nu(q_\epsilon z).
\eeq

On $M_2$ we then have (where we match at $t=T$)
\begin{align}
\phi_{2}(\tau_{2},\vec{x},z) = \frac{2\pi i}{\Gamma(\nu)}& e^{Q(-i\beta(\tau_2+iT)/z^2)-Q(\hat T_1)}e^{in\xi}  z^{1+d/2-in\alpha\beta}\nonumber\\
&\times \int  \frac{d\omega}{2\pi} \frac{d^d k}{(2\pi)^d}\ e^{-\omega (\tau_2 + iT-i\hat t_1)  + i \vec k\cdot \vec x}
\theta(-q^2)\left(\frac{|q|}{2}\right)^\nu J_\nu(|q|z).
\end{align}

For $t_1>\hat t_1$, the correlator $K(t_1,\hat t_1)$ is essentially $\phi_1$ itself. Given the choice of $Q(T)$, we have
\beq
K_{age}(t_1,\hat t_1;\vec x)=\left(\frac{t_1}{\hat t_1}\right)^{in\alpha\beta /2} K_{Schr}(t_1,\hat t_1;\vec x)
\eeq
This result is the expected one --- it displays the time-translation non-invariant scaling form, with exponent given by $K=i\alpha n\beta/2$. We see that for real $\alpha$, this time dependence is a phase.\footnote{In the aging literature, many specific systems have been studied numerically for which no such phase is present. To be capable of seeing such a phase, one must at least have a complex order parameter. As an example, it is possible that such behavior could be seen in $p_{x}+ip_{y}$ superconductors.} It is only for $\alpha$ complex that the correlator corresponds to a relaxation process. For generic $\alpha$, the correlator `spirals in' towards the Sch\"odinger correlator at late times. 

Note that for this physical interpretation, one expects that ${\rm Im}\ \alpha>0$. This is related holographically to normalizability of the solutions. Indeed if one traces the solution back to $t=-T$, for ${\rm Im}\,\alpha <0$, the prefactor of the solution blows up as $T\to\infty$, but is innocuous for $\alpha$ in the upper half plane.

\section{Conclusions}\label{sec:conclusions}

We have discussed the holography of the aging group, its geometric realization and the relevant correlation
functions as implied by the aging/gravity duality. We hope that this approach will lead to a better understanding of aging phenomena as it carries the seed of potential applications to a large host of phenomena involving polymeric materials, spin glasses, ferromagnets and granular media. We believe that various questions concerning couplings to various sources could be readily answered in the holographic context. The holographic approach can also potentially allow to address questions of thermodynamics and higher point correlations.

There are a few interesting questions one might like to pursue. A natural one is the embedding of the solutions we considered here into string theory. Another problem that is largely suggested by the condensed matter literature is the computation of the correlation function for operators with different conformal dimensions. In the holographic setting this problem seems more difficult and it might involve the existence of some domain wall solution.

It is worth stressing that from the condensed matter point of view one of the most important problems would be the {\it determination} of the critical exponents. Although the practical details of this problem remain completely elusive, it is worth noting that string theory provides a framework for understanding the process through holographic RG flows. In fact, it is plausible to imagine holographically starting with an ultraviolet theory that in the infrared becomes one of the geometries we have discussed, and in such a case, the value of exponents would presumably be selected.

\vskip .5cm

{\bf \Large Acknowledgements}

\vskip .5cm

We would like to thank Tom\'as Andrade, Eduardo Fradkin, Michael Gutperle, Mokhtar Hassaine, Don Marolf, Nam Nguyen-Hoang, Andrei Parnachev, Michel Pleimling, Diana Vaman and Mike Weissman for discussions. After completing this work, Ref. \cite{Nakayama:2010xq}, which has some overlapping content, was brought to our attention,  R.G.L., D.M. and L.P-Z. are thankful to the Aspen Center for Physics for hospitality during the initial stages of this project. L.P-Z. is thankful to the Kavli Institute for Theoretical Physics (KITP) and J.I.J. is thankful to the Michigan Center for Theoretical Physics (MCTP) ``Young High Energy Theorist Visitors" program for their hospitality during the final stages of this project. J.I.J. is supported by a Fulbright-CONICYT fellowship. This work is partially supported by the U.S. Department of Energy under contracts DE-FG02-95ER40899, DE-FG05-92ER40677 and DE-FG02-91-ER40709.
\vskip 1cm

\appendix

\section{Killing Equations}\label{AgingMetrics}

In this appendix we provide further details about the calculations presented in section (\ref{Sec:Metrics}).  Recall that the Ansatz  \eqref{metric ansatz} we start with already respects the symmetries $P,K,D,N$, and the corresponding generators take the same form as in the standard Schr\"odinger geometry \eqref{Son-McGreevy} (with $\de=2$). Thus, the question we pose is whether we can choose the set of metric functions in such a way that the metric also admits a Killing vector isomorphic to $C$, but not an $H$ generator. In order to allow for the appearance of additional isometries (besides $P,K,D,N$), one is led to set
\begin{align}
f_{zz}(T) &= k^{2}\left(1 + h(T)\right)^{2},
\end{align}
where $h(T) = T\dot{f}(T)/f(T)\neq -1$ and $T=\beta t/z^{2}$ as in the body of the paper, and $k$ is an integration constant that can be set to $k=1$ by a simple redefinition of parameters. A systematic analysis of Killing's equations fixes a general would-be isometry generator $V$ to be of the form
\begin{equation}
V = c_{P}P + c_{K}K + c_{D}D + c_{N}N + c_{C}C + c_{H}H,
\end{equation}
where $c_{P}$, $c_{K}$, etc are constants, and in addition to $D,K,P,N$ we have
\begin{align}
C &=\frac12 tz\left(\frac{h+2}{h+1}\right)\partial_{z} + t^{2}\partial_{t}   + tx\,\partial_{x}
+\left[\frac{z^{2}T}{2f(h+1)}\left(-T\dot{f}_{zt} + 2(h+1)f_{zt} + T^{2}\dot{f}_{tt} - 2Thf_{tt}\right) \right.\nonumber\\
&\phantom{=}\left. \qquad\qquad\qquad\qquad\qquad\qquad\qquad\qquad
+ \frac{k^{2}z^{2}}{4f}\left(-T\dot{h} + (h+1)(h+2)\right) + \frac{x^{2}}{2}\right]\partial_{\xi}\\
H&= \partial_{t} + \frac{z}{2t}\frac{h}{h+1}\partial_{z} + \left[\frac{\beta}{2tf(h+1)}\left(-T\dot{f}_{zt} + 2hf_{zt}
+T^{2}\dot{f}_{tt} - 2Thf_{tt}\right)\right. \nonumber\\
&\phantom{=}\left. \qquad\qquad\qquad\qquad\quad - \frac{\beta k^{2}}{4tTf}\left(T\dot{h} - h(h+1)\right)\right]\partial_{\xi}\, .
\end{align}
The crucial point is as follows: for a generator $V$ of this form, there is still one component of Killing's equations that remains to be satisfied, which is\footnote{Writing Killing's equations in the $(T,t,\xi,x)$ coordinate system, the remaining equation is the $Tt$ component.}
\begin{equation}\label{app: the last Killing eq}
0 =\left(\frac{c_{H}}{t^{2}} + c_{C}\right)G(T)\, ,
\end{equation}
where $G(T)$ is a function of $T$ only, given by
\begin{align}
G(T) &=+2{T}^{4}(h+1)\ddot{f}_{tt}-2{T}^{4}\dot{h}  \dot{f}_{tt} -2{T}^{3}h \dot{f}_{tt}+4{T}^{3}\dot{f}_{tt} -6{T}^{3} h^{2}\dot{f}_{tt} -4{T}^{3}f_{tt}\dot{h}   +4{T}^{2}  h^{3}f_{tt}   -4{T}^{2}f_{tt} h\nonumber\\
 &\phantom{=}-2{T}^{3}(h+1)\ddot{f}_{zt}+2{T}^{3} \dot{h} \dot{f}_{zt} +6{T}^{2} h(h+1)\dot{f}_{zt}
 -2f_{zt}Th(h+1)(2h+1)
  +2{T}^{2}f_{zt}\dot{h}  \nonumber\\
 &\phantom{=}-k^2h(h+1)^3(h+2)+k^2T\dot h(h+1)^2(3h+2)-{k}^{2}{T}^{2} (h+1)^2\ddot{h}\, .
\end{align}
Thus, in order to satisfy \eqref{app: the last Killing eq} we need either \{$c_{H} = 0$ and  $c_{C} = 0$\}, or  $\{G(T) = 0\}$. If $c_{H} =c_{C} = 0$ we are back to our starting Ansatz, that is, only the $4$ isometries $P,K,D,N$. So the only interesting case is when $G(T) = 0$. But this leaves $c_{H}$ and $c_{C}$ arbitrary, introducing 2 new isometries at the same time ($C$ and $H$). The remaining constraint $G(T) = 0$ should be thought of as an equation determining one of  $f_{tt},f_{zt},h$ (i.e. $f$) when the other two are given.
 Thus, our metrics are a family parameterized by 2 independent functions, that here we will choose to be $f$ and  $f_{tt}$. Remarkably, the equation $G(T) = 0$ can be integrated, providing the relation
\begin{equation}\label{expression for fzt}
f_{zt}(T) = T\left(f_{tt} + cf^{2}\right) - \frac{k^{2}}{4T}h(h+2)\, ,
\end{equation}
where $c$ is an arbitrary constant. This in turn simplifies the form of the generators we gave above:
\begin{align}
C &=\frac12 tz\left(\frac{h+2}{h+1}\right)\partial_{z} + t^{2}\partial_{t}   + tx\partial_{x}+\frac{1}{2}\left[x^{2} + \frac{z^{2}T^{2}}{f(h+1)}\left( \frac{k^{2}}{4T^{2}}\left(h+2\right)^{2}  +f_{tt}+cf^2\right) \right]\partial_{\xi}\\
H&= \partial_{t} + \frac{z}{2t}\frac{h}{h+1}\partial_{z} -\frac{\beta^{2}}{2z^{2}f(h+1)}\left(\frac{k^{2}}{4T^{2}} h^{2}+ f_{tt} + c f^2\right)\partial_{\xi}\, .
\end{align}


\providecommand{\href}[2]{#2}\begingroup\raggedright\endgroup

\end{document}